\documentclass[conference]{IEEEtran}
\IEEEoverridecommandlockouts
\usepackage{cite}
\usepackage{amsmath,amssymb,amsfonts}
\usepackage{algorithmic}
\usepackage{pstricks}
\usepackage{graphicx}
\usepackage{textcomp}
\usepackage{xcolor}
\usepackage{multirow}
\usepackage{bbm}
\usepackage{dsfont}
\usepackage{comment}
\usepackage{breqn}
\usepackage{amsmath}
\usepackage[linesnumbered,ruled,vlined]{algorithm2e}
\usepackage{caption}
\usepackage{subcaption}
\usepackage{multicol}
\usepackage{eso-pic}
\newcommand\AtPageUpperMyright[1]{\AtPageUpperLeft{%
 \put(\LenToUnit{0.5\paperwidth},\LenToUnit{-1cm}){%
     \parbox{0.5\textwidth}{\raggedleft\fontsize{9}{11}\selectfont #1}}%
 }}%
\newcommand{\conf}[1]{%
\AddToShipoutPictureBG*{%
\AtPageUpperMyright{#1}
}
}
\usepackage{mathrsfs}
\usepackage[scr=rsfs,cal=boondox]{mathalfa}
\newcommand{\nonl}{\renewcommand{\nl}{\let\nl\oldnl}}
\def\BibTeX{{\rm B\kern-.05em{\sc i\kern-.025em b}\kern-.08em
    T\kern-.1667em\lower.7ex\hbox{E}\kern-.125emX}}
\begin{document}

\title{Reconfigurable Low-latency Memory System for Sparse Matricized Tensor Times Khatri-Rao Product on FPGA}

\author{
    \IEEEauthorblockN{Sasindu Wijeratne\IEEEauthorrefmark{1}, Rajgopal Kannan\IEEEauthorrefmark{2}, Viktor Prasanna\IEEEauthorrefmark{1}}
    \IEEEauthorblockA{\IEEEauthorrefmark{1}Department of Electrical and Computer Engineering, University of Southern California, Los Angeles, USA}
    \IEEEauthorblockA{\IEEEauthorrefmark{2}US Army Research Lab, Los Angeles, USA}
    Email: kangaram@usc.edu, rajgopal.kannan.civ@mail.mil, prasanna@usc.edu
}

\maketitle

\begin{abstract}
Tensor decomposition has become an essential tool in many applications in various domains, including machine learning. Sparse Matricized Tensor Times Khatri-Rao Product (MTTKRP) is one of the most computationally expensive kernels in tensor computations. Despite having significant computational parallelism, MTTKRP is a challenging kernel to optimize due to its irregular memory access characteristics. This paper focuses on a multi-faceted memory system, which explores the spatial and temporal locality of the data structures of MTTKRP. Further, users can reconfigure our design depending on the behavior of the compute units used in the FPGA accelerator. Our system efficiently accesses all the MTTKRP data structures while reducing the total memory access time, using a distributed cache and Direct Memory Access (DMA) subsystem. Moreover, our work improves the memory access time by 3.5$\times$ compared with commercial memory controller IPs. Also, our system shows 2$\times$ and 1.26$\times$ speedups compared with cache-only and DMA-only memory systems, respectively.
\end{abstract}


\conf{2021 IEEE High Performance Extreme Computing Conference}

\section{Introduction}

Tensors are the de-facto representation of high-dimensional data. They have become the center for Machine Learning techniques\cite{7891546} such as recommender systems\cite{chen2021deep}\cite{kolda2020stochastic} and neural networks\cite{mondelli2019connection}\cite{cheng2020novel}. Tensors are also used in other domains including network analysis\cite{fernandes2020tensor}, chemistry\cite{taguchi2019drug}, and signal processing\cite{wen2020tensor}.

With the recent surge in Machine Learning, the attention towards tensor decomposition grew exponentially. Canonical polyadic decomposition (CPD)\cite{hong2020generalized} is one of the popular methods for tensor decomposition. CPD approximates the tensor as a sum of rank-one tensors. The alternating least squares method (CP-ALS) is the most common algorithm which use to compute CPD. CP-ALS computes a new factor matrix for each mode in each iteration. Matricized tensor by Khatri-Rao product (MTTKRP), kernel is the common bottleneck in this computation.

Since the sparsity of the real-world tensors\cite{10.1145/1345448.1345459}\cite{10.5555/2898607.2898816} is considerably high, specialized hardware accelerators are becoming common means of improving compute efficiency of tensor computations. But external memory access time has become the bottleneck due to irregular data access patterns.

There have been several techniques proposed in the literature to overcome the memory access time while accessing data with irregular access patterns\cite{10.1145/977091.977115}\cite{volos2016an}\cite{wijeratne2021programmable}. Caches\cite{5695314} are very productive in this regard if the required data fits in the cache and the data has spatial and temporal locality\cite{10.1145/106972.106981}. The ongoing approach for solving long memory access delays is to use onboard Block RAM (BRAM) in the FPGA as a data cache and facilitate data retrieval. An alternative solution is to look at multiple memory requests DMA transfers\cite{wijeratne2021programmable}\cite{10.1145/3020078.3021743}.

In this paper, we propose a memory system to significantly reduce the total memory access time on sparse tensors while performing MTTKRP. We also analyze the memory access patterns of the data structures used in the sparse-MTTKPR operation and suggest the best memory components to use with them. To the best of our knowledge, no prior work has proposed a memory system for sparse MTTKRP compute fabrics while analyzing the memory access patterns of its data structures.

The key contributions of this paper are:
 \begin{itemize} 
     
     \item \textbf{Analyzing Memory access patterns of sparse MTTKRP}: We analyze the memory access patterns of different data structures used in sparse MTTKPR computation. Then we propose memory modules (e.g., DMA controller and cache) that can use with the data structures to befit from their memory access patterns.

    \item\textbf{Reconfigurable memory system}: We propose a Local Memory Block (LMB) based memory system that can reconfigure depending on the targeted compute fabric and data layout used in the accelerator. We also experiment on different types of MTTKRP Processing Elements (PEs) and different memory system configurations, respectively.

    \item We evaluate the system on a Xilinx Alveo U250 board. It shows 3.5$\times$, 2$\times$, and 1.26$\times$ speedups in memory access time compared with commercial memory controller IPs, cache-only, and DMA-only memory systems.
     
 \end{itemize}

The rest of the paper is organized as follows, Sections II and III focus on understanding the sparse MTTKRP operation and prior work. Section IV discusses the architecture of the proposed memory design in-depth, while Section V presents the evaluation results. Finally, the paper is concluded in Section VI.

\section{Background}\label{background}
In this section, we first describe the notations used in the paper. Then we investigate the basics and the usage of MTTKRP kernel. 


\subsection{Notation}
This paper follows the definitions and symbols used in Tensor Algebra Compiler (TACO)\cite{8661185}.  The only difference is, we use capital calligraphic letters (e.g., $\mathscr{B}$) to represent tensors with N dimensions. 

Table \ref{notation} summarizes the symbols commonly used in the paper.

\begin{table}[ht!] 
\centering
\caption{Notation}
\label{notation}
\begin{tabular}{c|c}\hline
\textbf{Notation} & \textbf{Description}
\\\hline\hline
$\mathscr{B}$    &	Tensor	\\\hline
A  / C / D   &	Matrix	\\\hline
$\mathscr{B}_{i,j,k}$	&	Element at (i, j, k) of $\mathscr{B}$	\\\hline
A$_{i,j}$	&	Element at index (i, j) of A	\\\hline
nnz	&	Number of nonzeros in $\mathscr{B}$	\\\hline

\end{tabular}
\end{table}

\subsection{Use Cases of MTTKRP}

MTTKRP is the core computation kernel in the alternating least square (ALS) method, which is a popular method for finding the factor matrices in CPD. The CPD, with R as the decomposition rank, approximates tensor ($\mathscr{B}$) as the sum of R rank-one tensors: $\mathscr{B} \approx \sum_{r=0}^{R-1} a_r \circ d_r \circ c_r$, where $\circ$ denotes the outer product, and a, c, d are the components of the factor matrices A, C, and D. ALS method is a tensor approximation method computes using the above equation. In ALS, one of the factor matrices is updated at a time while fixing the rest. Algorithm \ref{ALS_algo} shows the steps of CP-ALS algorithm. The MTTKRP operation, which is the focus of this paper, takes place in lines 2, 3, and 4 of Algorithm \ref{ALS_algo} to compute each of the factor matrices of all modes.

\begin{algorithm}
\DontPrintSemicolon
\KwIn{$\mathscr{B}$ $\in$ $\mathbb{R}^{I\times J\times K}$}
\KwOut{A $\in \mathbb{R}^{I\times R}$ , C $\in \mathbb{R}^{K\times R}$, D $\in \mathbb{R}^{J\times R}$ }
\While{no improvement or maximum iterations reached}{
A $\leftarrow  \mathscr{B}_{(1)} (D \odot C)(C^TC*D^TD)$ \;

D $\leftarrow  \mathscr{B}_{(2)} (A \odot C)(C^TC*A^TA)$ \;

C $\leftarrow  \mathscr{B}_{(3)} (D \odot A)(A^TA*D^TD)$ \;

Normalize columns of A,B,C and store in $\lambda$
}
return A, B, C

\caption{{\sc CPD-ALS for third-order tensors}}
\label{ALS_algo}
\end{algorithm}

\subsection{MTTKRP}
MTTKRP consists of N-dimensional tensor multiply with N-1 factor matrices. Here, N is the order of the tensor. Mathematically, mode-1 MTTKRP (for three-dimensional tensors) can be expressed as Equation \ref{mttkrp_1}. In this equation, A, C, and D are dense matrices while $\mathscr{B}$ is a three-dimensional tensor (matricizied along with the first mode), and $\odot$ denotes the Khatri-Rao product.

\begin{dmath}
A =  \mathscr{B}_{(1)} (D \odot C)
\label{mttkrp_1}
\end{dmath}

This operation can also express in index notation as Equation \ref{mttkrp_2}.

\begin{dmath}
A_{i,r} = \sum_{j=0}^{J-1} \sum_{k=0}^{K-1} \mathscr{B}_{i,j,k} \cdot D_{j,r} \cdot C_{k,r}
\label{mttkrp_2}
\end{dmath}

In the following sections, we target 3D tensors since they are easy to understand. Even though we focus on 3-dimensional tensors as examples in this paper, we can expand our work into higher dimensions without any extra work. 

The sparsity of the tensors creates irregular data access patterns, which makes the external memory access more challenging than the dense MTTKRP. All the real-world tensors are sparse in nature. Algorithm \ref{gpu_paper_ref_mttkpr} shows the sequential sparse MTTKRP (spMTTKRP) approach for third-order tensors in COO format\cite{TTB_Sparse}\cite{8821030}.

\begin{algorithm}
\DontPrintSemicolon
\KwIn{indI[nnz], indJ[nnz], indK[nnz], vals[nnz], D[J][R], C[K][R]}
\KwOut{A[I][R] }
\For{z = 0 to nnz}{
i = indI[\textit{z}] \;
j = indJ[\textit{z}] \;
k = indK[\textit{z}] \;

\For{r = 0 to R}{
    A[i][r] += vals[\textit{z}] $\cdot$ D[j][r] $\cdot$ C[k][r] \;
}
}
return A

\caption{{\sc COO based spMTTKRP for third order tensors}}
\label{gpu_paper_ref_mttkpr}
\end{algorithm}

We use the term \textit{fibers} throughout this paper. \textit{Fibers} are the building blocks of tensors. Fiber is a one-dimensional fragment of a tensor obtained by fixing all indices but one. Tensor fibers are a higher-order extension of matrix rows and columns. The fibers of a three-dimensional tensor can be represented as $\mathscr{B}_{:,j,k}$, $\mathscr{B}_{i,:,k}$, and $\mathscr{B}_{i,j,:}$. Similarly, for a matrix $C$, its rows $C_{i,:}$ and columns $C_{:,j}$ are its fibers. In Equation \ref{mttkrp_m1} and \ref{mttkrp_m2}, the term fiber corresponds to the rows of the factor matrices. 

\begin{figure*}[t] 
\centering
\includegraphics[width=\textwidth]{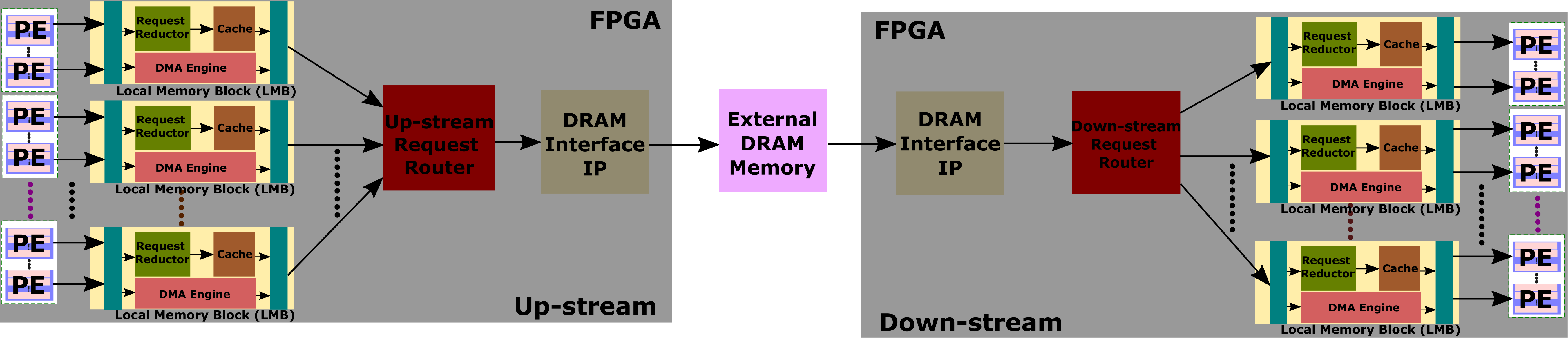}
\caption{Overall Architecture}
\label{overall}
\end{figure*}

\section{Related Work}

M. Asiatici et al.\cite{10.1145/3289602.3293901}\cite{8892073} present miss optimized cache system for FPGA accelerators. The key idea of their work is to increase the reusability to serve multiple requests from the compute fabric on the fly without relying on long-term storage in cache. They also use cuckoo hash table-based MSHR techniques to avoid the memory traffic due to secondary cache misses. In our work, instead of having only a cache-based system, we propose a cache $+$ DMA system to accelerate the external memory accesses. Instead of using MSHR, we introduce an XOR hash-table-based Recent Request Status Holder (RRSH) unit closer to the compute fabric. Unlike MSHR, RRSH takes care of secondary cache misses while reducing the memory traffic between the cache and compute units. Additionally, we want to emphasize that the focus of our work is entirely on the MTTKRP kernel.  

S. Aananthakrishnan et al.\cite{ DBLP:journals/corr/abs-2010-06277} have proposed a memory-optimized large-scale graph processor on ASICs. This paper emphasizes the importance of parallel support for bulk transfers and cache transfers for graph workloads. Even though we are focusing on a different domain of algorithms and proposing an entirely different memory system, we still observed the importance of supporting cache and DMA transfers simultaneously.

N. Srivastava et al.\cite{8735529}\cite{ 9065579} have implemented a tensor computation library for FPGAs and CGRAs. These papers describe compute fabrics to accelerate spMTTKRP using novel algorithmic optimizations. In our paper, we explore the memory optimizations we can implement using a multi-faceted approach. The spMTTKRP algorithms describe in\cite{8735529}\cite{ 9065579} can be used as the compute fabrics to our proposed memory system.

\section{Memory System Architecture}

We develop a unified memory system to share among multiple PEs that process MTTKRP. The state-of-the-art MTTKRP compute accelerator designs execute either Equation \ref{mttkrp_m1} or Equation \ref{mttkrp_m2} in their compute fabrics. Here, fiber$_{out}$ represent one or more output fibers of the output matrix, fiber$_j$ and fiber$_k$ represent a single fiber from input matrices, scalar represents a scalar value from high dimensional input tensor, and $\circ$ is corresponding to the Hadamard product of the two input vectors. Even though the internal construction of PEs is different in different state-of-the-art accelerators, they follow the same memory access pattern. The types of memory accesses can be summarized as follows: (a) load the input fibers of the matrices from the external memory, (b) load the scalar of the input tensor from the external memory, (c) store the output fiber to the external memory. In a parallel and pipeline system,  (a), (b), and (c) occurs in parallel for multiple fibers. Favorably, our memory system can receive the data from each step simultaneously.

\begin{dmath}
fiber_{out} = scalar \cdot \sum_{K}\sum_{J} (fiber_{k} \circ fiber_{j})
\label{mttkrp_m1}
\end{dmath}

\begin{dmath}
fiber_{out} = \sum_{K}\sum_{J} fiber_{k} \circ (scalar \cdot fiber_{j})
\label{mttkrp_m2}
\end{dmath}

Our proposed architecture supports 2 types of memory transfers:
\begin{enumerate}
  \item \textbf{cache-line transfers}: Nurtures single memory accesses. Load/store individual requests in minimum latency. The access patterns with high spatial and temporal locality are accessed using cache-line transfers. The element-wise access of scalar values from the input tensor shows spatial and temporal locality. Therefore, we use cache-lines to load the scalars from external memory.
  \item \textbf{DMA transfers}: Supports streaming accesses. Load/store operations on all requested data with minimum latency from memory. Loading and storing matrix fibers is a streaming memory access process. Therefore, we use the DMA engine to transfer matrix fibers between FPGA and external memory.
\end{enumerate}

Figure \ref{overall} shows the overall architecture of the proposed memory system. The upstream logic designates the data path from PEs to the external memory, while the downstream logic shows the data path from external memory to the PEs. The overall data path is symmetrical over the external memory.

\begin{figure}
\centering
\includegraphics[width=\linewidth]{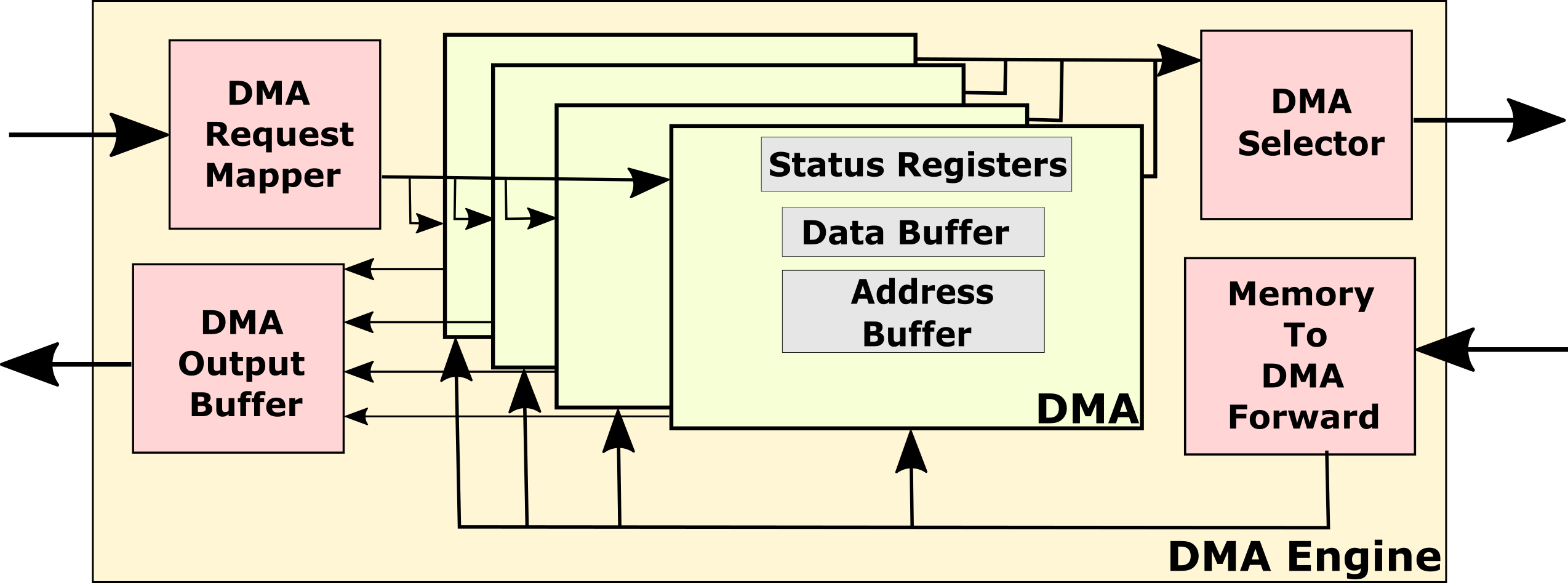}
\caption{DMA Engine}
\label{dma}
\end{figure}

The Local Memory Blocks (LMBs) are the basic building blocks of our proposed memory system. A LMB has a Request Reductor, non-blocking cache, and a DMA Engine.  Each LMB connects to one or more PEs. The complexity of the connection between PEs and LMB exponentially increases with the number of PEs connected to a single LMB. As shown in Figure \ref{overall}, we use multiple LMB distributed among all the PEs to avoid such unnecessary complexity. Using more than one LMB does not impact the memory consistency model of the MTTKRP accelerators. The reasons behind maintaining the consistency are: (a) Only the PEs connected to the same LMB update the same output fiber, (b) The input fibers or scalars do not update during the same MTTKRP operation.

\subsection{DMA Engine}


The DMA engine is in charge of communicating the fibers of the matrices between PEs and the external memory.  We store the matrices in row-major order because the MTTKRP algorithm encourages row-wise matrix accesses. The matrices in row-major order make the fiber accesses into a bulk of sequential memory transfers. Figure \ref{dma} shows the DMA engine and its internals. It has several DMA buffers inside. Therefore, it can support multiple fiber reads and writes simultaneously. The number of DMA buffers is proportional to the number of PEs connected to the same LMB. A large number of DMA buffers in a LMB can reduce the maximum operating clock frequency due to increased hardware routing complexities.

\subsection{Cache}

The cache focuses on satisfying a single memory request with minimum latency. Within the cache engine, we explore the spatial and temporal locality. It loads the scalars of the tensor from the external memory.

\begin{figure}
\centering
\includegraphics[width=\linewidth]{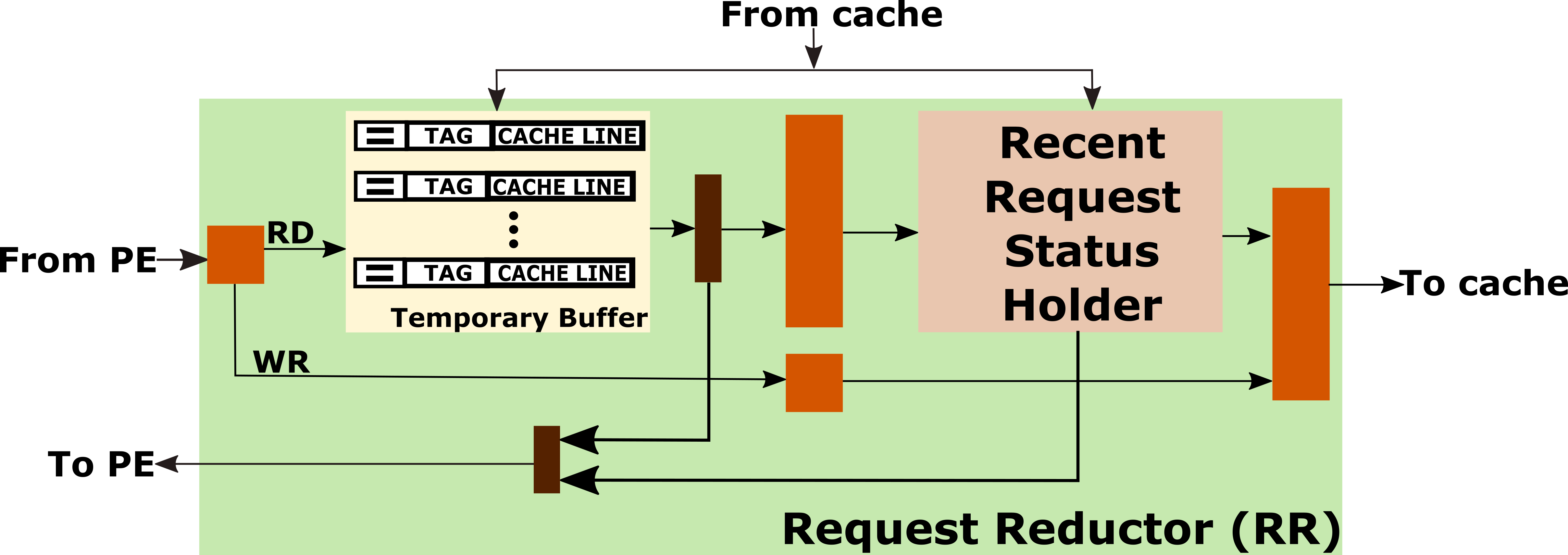}
\caption{Internals of Request Reductor}
\label{reductor}
\end{figure}

Our non-blocking cache uses a 3-stage pipeline to achieve high frequency. We keep the cache-line width similar to the data width of DRAM Interface IP to avoid implementation complexities. Instead of forwarding a single element from the cache to PEs, the cache passes the complete cache-line to the Request Reductor (RR). Then, the RR advances the requested portion to each PE while storing the incoming cache-line inside its temporary buffer.

\subsection{Request Reductor (RR)}

\begin{algorithm}
\DontPrintSemicolon
Let PE$_{v}$ denote the v$^{th}$ PE on FPGA, $0 \leq v < p$ \;
Let Partition$_{q}$ denote the q$^{th}$ partition, $0 \leq i < p$ \;
\KwIn{NNZ[M] = \{indI[M], indJ[M], indK[M], vals[M]\}, B[J][R], C[K][R]}
\KwOut{Y[I][R]}
\While{not done}{
\For{each partition${_q}$ \textbf{parallel}}{
\For{z = 0 to M/p}{
i = indI[z] \;
j = indJ[z] \;
k = indK[z] \;
\If{current$_I$ $\neq$ indI}{
\For{r = 0 to R}{
Y[current$_I$][r] = temp$\_$Y[r] \;
}

current$_I$ = indI \;
\For{r = 0 to R}{
temp\_Y[r] = vals[z]$\times$B[j][r]$\times$C[k][r] \;

}

} 
\Else{
\For{r = 0 to R}{
    temp\_Y[r] += vals[z]$\times$B[j][r]$\times$C[k][r] \;
    
}
}
}
}
}

\caption{{\sc Parallel-MTTKRP Algorithm}}
\label{alg:para_mttkrp_my}
\end{algorithm}

RR converts element-wise cache reads to cache-line accesses. As shown in Figure \ref{reductor}, the RR is a 2-stage pipeline. In the first step, a temporary buffer stores the most recent memory reads. It is a CAM-based memory implementation that keeps the most recent external memory reads inside. Since CAMs are hardware expensive, we keep the number of elements in the buffer small.

\begin{table*}
\centering
\caption{Module Configuration and Resource Utilization}
\label{m_config}
\begin{tabular}{|c|c|c|c|c|c|}\hline
\multicolumn{1}{|c|}{Module} & \multicolumn{1}{c|}{Specification} & \multicolumn{4}{c|}{Resource Utilization}\\
\cline{3-6}
\multicolumn{1}{|c|}{Name} &  & LUT(\%) & FF(\%) & BRAM(\%) & URAM(\%) \\
\hline\hline
   &  \textbf{Configuration-A}  & & & & \\\hline\hline
   & Degree of set-associativity = 2    & & & & \\
Cache   & No. of cache-lines = 8192   & 1.87 & 1.24 & 0.24 & 1.25 \\
   & Cache-line width  = 512    & & & & \\\hline

DMA   & No. parallel DMA supported = 4    & & & & \\
Engine   & Size of single DMA buffer = 256 B   & 0.04 & 0.01 & - & 0.25 \\\hline

Request   & No. of entries in RRSH = 4096    & & & & \\
Reductor   & No. of entries in Tempory Buffer = 8   & 0.08 & 0.10 & - & 1.25 \\\hline

   & Includes a cache, a DMA Engine, & & & & \\
LMB   & and a Request Reductor   & 2.03 & 1.41 & 0.24 & 2.75 \\\hline

\textbf{Complete}   &    & & & & \\
\textbf{System}   & \textbf{No. of LMBs = 1}   & \textbf{2.25} & \textbf{1.54} & \textbf{0.24} & \textbf{2.75} \\\hline\hline
   &  \textbf{Configuration-B}  & & & & \\\hline\hline
   & Degree of set-associativity = 1    & & & & \\
Cache   & No. of cache-lines = 4096   & 0.65 & 0.64 & 0.06 & 0.63 \\
   & Cache-line width  = 512    & & & & \\\hline

DMA   & No. parallel DMA supported = 4    & & & & \\
Engine   & Size of single DMA buffer = 256 B   & 0.04 & 0.01 & - & 0.25 \\\hline

Request   & No. of entries in RRSH = 4096    & & & & \\
Reductor   & No. of entries in Tempory Buffer = 8   & 0.08 & 0.10 & - & 1.25 \\\hline

   & Includes a cache, a DMA Engine,    & & & & \\
LMB   & and a Request Reductor   & 0.85 & 0.81 & 0.06 & 2.13 \\\hline

\textbf{Complete}   &    & & & & \\
\textbf{System}   & \textbf{No. of LMBs = 4}   & \textbf{3.61} & \textbf{3.35} & \textbf{0.24} & \textbf{8.52} \\\hline
\end{tabular}
\end{table*}

If requested data is not in the temporary buffer, the read request advances to the Recent Request Status Holder (RRSH). RRSH keeps the status of recently forwarded requests to the cache. If the incoming read request belongs to one of the pending cache-line requests, the PE id and address are kept in the RRSH. When a cache-reply from cache reaches the RRSH, the pending requests corresponding to that cache line are satisfied by sending the corresponding data elements to the requested PEs. RRSH helps to convert element-wise cache access to cache-line-wise accesses. It drastically reduces the memory traffic to the cache. RRSH logic can be implemented using a hash table. In our work, we use XOR-base hash table\cite{xor_hash} considering its high throughput and scalability.

 \subsubsection{XOR-based Hash Tables}
R. Zhang et al.\cite{9286199} propose XOR-based hash tables that can support multiple parallel pipelines with high throughput. For stall-free execution, our work requires 2 PE versions of the hash table. Each PE connects to the cache interface and PE side separately. The total number of entries in the hash table is proportional to $\frac{total\; number\; of\; entries\; in\; the\; local\; cache}{local\; cache\; associativity}$. The width of a table is proportional to $ (Tag\; width\; +\; Number\; of\; PEs)$ and it connects with $(RR\; \times \; number\; of\; data\; elements\; per\; cache-line)$.

\subsection{Request Routers}

The request router has 2 responsibilities. They are (a) receive memory requests from different LMB units and forward them to the DRAM interface IP, (b) forward the data coming from external memory to the LMB units.

\subsection{Reconfigurability of the Overall Design}
Users can configure our design during the synthesis step. The configuration heavily depends on the characteristics of the compute fabric and data layout. In the current literature, all the FPGA or CGRA based implementations use a variation of COO Format (e.g., CISS) to store the high dimensional tensors. Meanwhile, the dense matrices use a row-major format. We assume all the compute fabrics that use our memory system use the same kind of memory layout. Under these conditions:
 (a) the sparse tensors use the cache due to the spatial and temporal locality of the data access pattern, (b) The fibers of the matrices are accessed through DMAs due to their high spatial locality and poor temporal locality.

In Section \ref{config_sys}, we dive deeper into configuring the number of LMBs depending on the type of the compute fabric. Our experiments show that the performance improvement due to the total number of DMAs in an LMB saturates after 4 DMAs. Increasing the number of DMAs also negatively impacts the maximum operating frequency due to increased place and route complexity. We further observed that the cache size also influences the maximum operating frequency of the overall design. We keep cache size including, the total number of cache lines and cache line width, as configurable parameters. We maintain the cache line width equal to the memory interface IP’s data width since it avoids design complexities of the cache.

\section{Evaluation}

\subsection{Experiment Methodology}
We implement different configurations of the memory controller on Xilinx Alveo U250 FPGA\cite{xilinxalveo} using Verilog HDL. This device has 1,728 K LUTs, 3,456 K Flip-flops, and 327 MB of URAM memory. Simulations, synthesis, and place-and-route implementations are performed using Xilinx Vivado Design Suite 2020.2. In our work, we focus on accessing a single external DRAM component efficiently. We use the Xilinx memory interface IP\cite{xilinxddr} to connect our design with the external memory. For the Alveo U250 board, Xilinx provides Memory Interface IP with 31-bit address width and 512-bit data width (with ECC turned on).

\subsubsection{Datasets}
Table \ref{datasets} summarizes the characteristics of the synthetic 3D tensors we use in our experiments. Most of the state-of-the-art accelerators use a variation of the COO format to store data. For example, N. Srivastava et al.\cite{8735529}\cite{ 9065579} use Compressed Interleaved Sparse Slice (CISS) format, which is also a variation of COO format. In this section, the tensors use variations of the COO format. Each element in the tensor includes its coordinate vector following the value. The total size of one 3D tensor element is 16 Bytes. We use 32 bits to store each coordinate and value. The dense matrices are stored in row-major order while keeping each element 4 Byte. We set the number of elements in a row of a matrix to 32.

\begin{table}[ht!] 
\centering
\caption{Sparse 3D Tensor Datasets}
\label{datasets}
\begin{tabular}{|c|c|c|c|c|}\hline
\textbf{Tensor} & \textbf{Dimensions} & \textbf{Nonzeros} & \textbf{Density}
\\\hline\hline
Synth_01 & 22K $\times$ 22K $\times$ 23M & 28M & 2.37E-09
\\\hline
Synth_02 & 3M $\times$ 2M $\times$ 25M & 144M & 9.05E-13
\\\hline
\end{tabular}
\end{table}

\subsection{Baselines}
We compare our proposed memory system with 3 approaches.

 \subsubsection{Memory Controller IP Only}: Commercial memory interface IP\cite{xilinxddr}\cite{intelddr} directly connected to a compute fabric of a given accelerator. It is the naïve method of connecting the external memory.

    \subsubsection{Cache-only}:  Compute fabric of the accelerator connects to the external memory through a cache. This approach is identical to replacing the LMB in our system with the cache.

    \subsubsection {DMA-only}: All the requests of the accelerator are accessed through a DMA. This approach is identical to replacing the LMB in our system with DMAs. Here, a DMA engine can load/store a single DMA request at a time.

\subsection{Configurations of the Proposed System} \label{config_sys}
There are 2 types of spMTTKRP kernel accelerators based on the communication between external memory and compute fabric.
\\
\textbf{Type-1:} 
The systolic array-based spMTTKRP compute fabrics\cite{8735529}\cite{9065579} in the current literature have a single point of access to the external memory for each type of data structure. After loading the data to the compute fabric, they are shared among PEs using the PEs near the memory system. For instance,\cite{9065579} use shared Loading and Storing Units for each type of data structure (i.e., Matrix Loading Unit (MLU), Tensor Loading Unit (TLU), and Matrix Store Unit (MSU)) and share the data among PEs using the edge PEs.
\\
\textbf{Type-2:} 
The compute fabrics with multiple points of access to external memory fall into this category. For instance, Algorithm \ref{alg:para_mttkrp_my} can be mapped into a compute fabric with multiple PEs with independent memory accesses. Here, the computations executed by each PE are independent of others. The memory system should be sophisticated enough to handle this type of compute fabric.

In the experiments, we use 2 types of configurations depending on the above communication types between external memory and compute fabric.
\\
\textbf{Configuration-A:} For Type-1, having multiple LMBs does not help because their data structures only have a single point of access to the external memory. Therefore, with Type-1 compute fabrics, we use a single Large LMB. Table \ref{m_config} shows the resource utilization of each module in the memory system that we used in our experiments.
\\
\textbf{Configuration-B:} For Type-2, multiple LMBs can increase the performance. Their compute fabric has several points of access for each of the data structures. In our experiments, we use 4 LMBs, with each LMB connected to a PE. Table \ref{m_config} summarizes the resource utilization of each module in the design we used in our experiments.

\subsection{Performance}
\begin{figure}
\centering
\includegraphics[width=\linewidth]{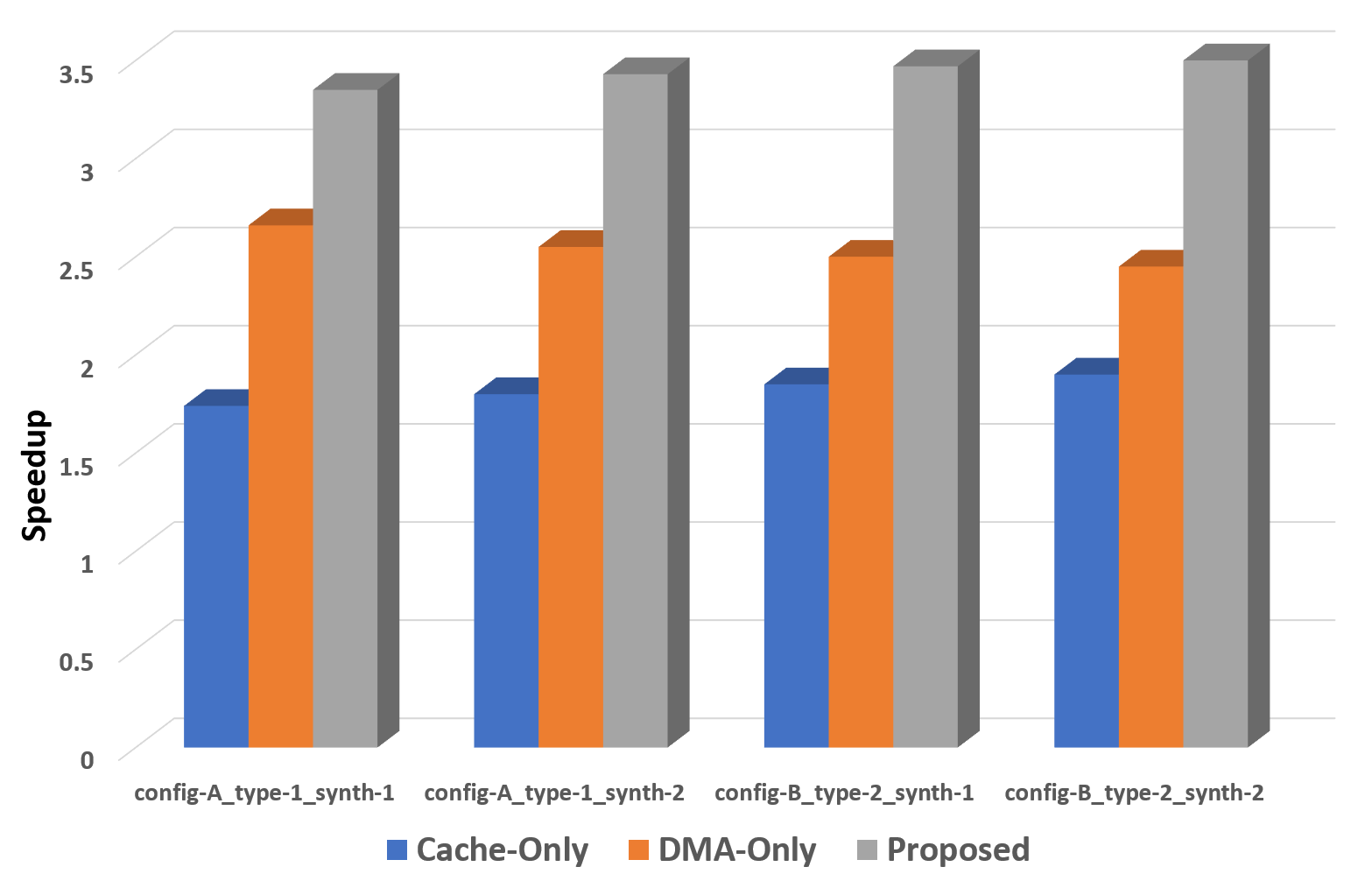}
\caption{Speedup with different memory systems over direct connection to commercial Xilinx memory  controller IP as the baseline. The naming convention of each category is ($<$configuration of proposed system$>\_<$type of the compute fabric$>\_<$dataset$>$). }
\label{results}
\end{figure}

As we discussed previously, we ran the experiments on different datasets with different memory system configurations. Figure \ref{results} shows the improvement after using our proposed memory systems over other alternatives. We compute the memory access speedup by comparing the total memory access time with the commercial memory controller IP-only setting. Our proposed system achieved around 3.5x speedup compared to the commercial memory controller IP-only setting. Also, compared with cache-only and DMA-only memory systems, our approach achieved around 2x and 1.26x speedup, respectively.

Our approach always achieves higher speedup compared to the cache-only setting. In MTTKRP, the consecutive secondary cache misses to the same cache line occur while accessing the fibers. In modern caches, secondary misses are avoided using Miss Status Holding Registers (MSHR). But conventional MSHR can not handle a large number of secondary cache misses without losing the performance. Further, the memory traffic between the cache and compute fabric can also reduce the performance in the cache-only setting. In our method, we reduce this traffic while avoiding secondary misses using the Request Reductor.

DMAs do not exploit the temporal locality of data. Further, there can be garbage data in DMA transactions when the length of the data requests is shorter than the width of the memory interface IP. Our proposed system outperforms the DMA-only setting simply because it avoids the above issues with DMAs.

\section{Conclusion}

In this paper, we propose a multifaceted memory system that reduces the total memory access time of MTTKRP on FPGAs. Our unified hardware can be reconfigured in the compile-time, depending on the orientation of the computing units and memory layout of tensors. The scalable Local Memory Blocks can handle the memory access pattern of MTTKRP efficiently while achieving 3.5$\times$ speedup compared to the commercial memory controller IPs. Also, our system shows 2$\times$ and 1.26$\times$ speedups compared with cache-only and DMA-only memory systems, respectively.

\section{Acknowledgment}
This work was supported by the U.S. National Science Foundation (NSF) under grant OAC-1911229 and CNS-2009057.

\bibliographystyle{IEEEtran}
\bibliography{hpec20}

\end{document}